\documentclass[prl,aps,showpacs,twocolumn,superscriptaddress,nofootinbib]{revtex4}
\usepackage{amsmath, amsthm, amssymb}
\usepackage{subfigure}
\usepackage{graphicx}
\usepackage{epstopdf}
\usepackage{pdfpages}
\usepackage{array}
\usepackage{multirow}
\usepackage{hyperref}
\usepackage{rotating}
\usepackage{placeins}
\usepackage{upgreek,xspace}
\def\textmu{\ensuremath\upmu}
\def\textalpha{\ensuremath\upalpha}
\def\textbeta{\ensuremath\upbeta}
\def\textgamma{\ensuremath\upgamma}

\usepackage{stackengine}
\usepackage[percent]{overpic}
\usepackage[symbol*]{footmisc}
\DefineFNsymbolsTM{otherfnsymbols}{%
  \textdagger    \dagger
  \textasteriskcentered *
  \textsection   \dagger
  \textbardbl    \|%
  \textparagraph \mathparagraph
  \textbullet \circ
  \textdaggerdbl \ddagger
}%
\setfnsymbol{otherfnsymbols}

\newcommand{\nuc}[2]{$^{#2}\rm #1$}

\newcommand{\fprompt}{F\ensuremath{_\mathrm{prompt}}\xspace}
\newcommand{\kevee}{keV$_{\mathrm{ee}}$}
\newcommand{\kevr}{keV$_{\mathrm{r}}$}

\begin{document}
\title{First results from the DEAP-3600 dark matter search with argon at SNOLAB}

\author{P.-A.~Amaudruz}
\affiliation{TRIUMF, Vancouver, British Columbia, V6T 2A3, Canada}

\author{M.~Baldwin}
\affiliation{Rutherford Appleton Laboratory, Harwell Oxford, Didcot OX11 0QX, United Kingdom}

\author{M.~Batygov}
\affiliation{Department of Physics and Astronomy, Laurentian University, Sudbury, Ontario, P3E 2C6, Canada}

\author{B.~Beltran}
\affiliation{Department of Physics, University of Alberta, Edmonton, Alberta, T6G 2R3, Canada}

\author{C.\,E.~Bina}
\affiliation{Department of Physics, University of Alberta, Edmonton, Alberta, T6G 2R3, Canada}

\author{D.~Bishop}
\affiliation{TRIUMF, Vancouver, British Columbia, V6T 2A3, Canada}

\author{J.~Bonatt}
\affiliation{Department of Physics, Engineering Physics, and Astronomy, Queen's University, Kingston, Ontario, K7L 3N6, Canada}

\author{G.~Boorman}
\affiliation{Royal Holloway University London, Egham Hill, Egham, Surrey TW20 0EX, United Kingdom}

\author{M.\,G.~Boulay}
\affiliation{Department of Physics, Carleton University, Ottawa, Ontario, K1S 5B6, Canada}
\affiliation{Department of Physics, Engineering Physics, and Astronomy, Queen's University, Kingston, Ontario, K7L 3N6, Canada}

\author{B.~Broerman}
\affiliation{Department of Physics, Engineering Physics, and Astronomy, Queen's University, Kingston, Ontario, K7L 3N6, Canada}

% ...

\collaboration{DEAP-3600 Collaboration}
\noaffiliation
\date{\today}
\begin{abstract}
This paper reports the first results of a direct dark matter search with the DEAP-3600 single-phase liquid argon (LAr) detector. 
The experiment was performed 2~km underground at SNOLAB (Sudbury, Canada) utilizing a large target mass, with the LAr target contained in a spherical acrylic vessel of 3600~kg capacity. The LAr is viewed by an array of PMTs, which would register scintillation light produced by rare nuclear recoil signals induced by dark matter particle scattering.
An analysis of 4.44 live days (fiducial exposure of 9.87~tonne$\cdot$days) of data taken during the initial filling phase demonstrates the best 
electronic recoil rejection using pulse-shape discrimination in argon, with leakage $<$1.2$\times$10$^{-7}$ (90\%~C.L.) between 15 and 31~\kevee.
No candidate signal events are observed, which results in the leading limit on WIMP-nucleon spin-independent cross section on argon, $<$1.2$\times$10$^{-44}$~cm$^2$ for a 100~GeV/c$^2$ WIMP mass (90\%~C.L.).
\end{abstract}
\pacs{95.35+d, 29.40.Mc, 26.65.+t, 34.50.Gb, 07.20.Mc, 12.60.Jv}
\maketitle
It is well established from astronomical observations that {\em dark matter} (DM) constitutes most of the matter in the Universe~\cite{planck}, accounting for 26.8\% of the energy density, compared to 4.9\% for ordinary matter. Weakly Interacting Massive Particles (WIMPs) are one of the leading DM candidates. 
Direct detection of WIMPs from the galactic halo is possible via elastic scattering, producing nuclear recoils (NR) of a few tens of keV.

This paper reports on the first DM search from DEAP-3600, a liquid argon (LAr) detector which uses single-phase technology, registering only the primary scintillation light from the target medium.  
This is the first DM search result from a LAr detector, of any technology, exceeding a 1~tonne target mass, and the first such result from a single-phase detector, of any target species, at this scale.
We emphasize the importance of exceeding the tonne scale: thus far only one technology, liquid Xe TPCs, has achieved 1 tonne fiducial mass while a credible direct detection discovery of DM will require observation in multiple target species.  
Further, while the WIMP mass reach of collider experiments is limited by beam energy, direct detection experiments are limited only by total exposure, and so a large enough underground detector with sufficiently low backgrounds can access high WIMP mass regions not accessible to colliders.  
The DEAP-3600 single-phase design offers excellent scalability to ktonne-scale LAr detectors~\cite{ichep,det}.

In this paper we report the best background rejection using pulse-shape discrimination (PSD) in argon at low energy threshold, most relevant for WIMP searches. 
The PSD uses the substantial difference in LAr scintillation timing between NR and electronic recoils (ER) to reject the dominant \textbeta/\textgamma\ backgrounds~\cite{psd0,psd} at the $10^{-7}$ level, 4 orders of magnitude beyond that achieved in LXe.  
This capability will enable a large underground detector using argon to reject the electron backgrounds from solar neutrinos and reach the neutrino floor defined by coherent scattering of atmospheric neutrinos.
Employing this PSD, this paper reports a background-free DM search in 9.87~tonne-day exposure, resulting in the best limit on the WIMP-nucleon cross section measured with argon, in the high WIMP mass regime second only to Xe TPC-based searches.

The detector is comprised of an atmospheric LAr target contained in an acrylic vessel (AV) cryostat capable of storing 3600~kg of argon. The AV is viewed by 255 Hamamatsu R5912-HQE photomultiplier tubes (PMTs) detecting scintillation light from the target. The PMTs are coupled to the AV by 50~cm-long acrylic light guides (LGs).
The inner AV surface was coated in-situ with a 3~\textmu m layer of wavelength shifter, 1,1,4,4-tetraphenyl-1,3-butadiene (TPB) to convert 128~nm Ar scintillation light into blue light transmitted through acrylic. 
The AV neck is wrapped with optical fibers read out by PMTs, to veto light emission in the AV neck region.
The detector is housed in a stainless steel spherical shell immersed in an
8~m diameter ultrapure water tank.
All detector materials were selected to achieve the background target of $<$0.6~events in a 3~tonne-year~\cite{det}.
To avoid \nuc{Rn}{222}/\nuc{Pb}{210} contamination of the AV surface, the inner 0.5~mm layer of acrylic was removed in-situ after construction; Rn exposure was then strictly limited.

PMT signals are decoupled
from the high voltage by a set of custom analog signal-conditioning boards, digitized (CAEN V1720) and handled by MIDAS DAQ~\cite{daq}.

The PMT charge response functions are calibrated daily with a system of 22 pulsed-LED-driven fibres injecting 435~nm light\cite{det}. A PMT charge response model is used to calculate the mean single photoelectron (SPE) charges, $\hat{\mu}_{SPE}$, with the combined 3\% statistical and systematic uncertainty~\cite{pmt}.
 A Monte Carlo model of the detector, using the \textsc{Geant4}-based RAT~\cite{rat}, includes a full PMT signal simulation based on in-situ measured time vs.~charge distributions for noise sources: late, double, and after-pulsing (AP) for each PMT~\cite{ap,pmt,det}.

The charge of each identified pulse is divided by the PMT-specific $\hat{\mu}_{SPE}$ to extract the number of photoelectrons (PEs). \fprompt\ is defined for each event as the ratio of prompt to total charge,
$\mathrm{\fprompt} \equiv \frac{\sum_{\{i|t_i\in(-28~\mathrm{ns}, 150~\mathrm{ns})\}}Q_i}{\sum_{\{i|t_i\in(-28~\mathrm{ns}, 10~\mathrm{\textmu s})\}}Q_i}$,
where $Q_i$ is the pulse charge in PE and $t_i$ is the pulse time relative to the event time. The relative timing of each channel is calibrated with a fast laser source; the resulting overall time resolution is 1.0~ns. \fprompt\ is a powerful discriminator because it is sensitive to the ratio of excited singlet to triplet states in LAr, $\frac{I_1}{I_3}$, with lifetimes of 6 and 1300~ns~\cite{heindl}, respectively. This ratio is significantly different for ER and NRs.

The detector trigger was designed to accept all low-energy events above threshold, all high-\fprompt\ NRs and to cope with approximately 1~Bq/kg \nuc{Ar}{39} activity of LAr~\cite{ar39}. The PMTs signal is continuously integrated in windows 177~ns and 3100~ns wide, from which the prompt energy (E$_\mathrm{trigger}$) and ratio of prompt and wide energies are calculated. All NR-like triggers with E$_\mathrm{trigger}>$40~PE, but only 1\% of \nuc{Ar}{39}-decay-like triggers, are digitized; summary information is recorded for all events.
For NR-like events above the analysis PE threshold, the trigger efficiency in the experiment livetime is measured to be (100$^{+0.0}_{-0.1}$)\%, by running in a very low threshold mode and after low-level cuts removing pile-up (Table~S1 in Supplemental Material~[]). For ER-like events the measured trigger efficiency is $<$100\% below 120~PE because of their lower prompt charge.

Stability of the LAr triplet lifetime, $\tau_3$, was verified with a fit accounting for dark noise, TPB fluorescence~\cite{segreto}, and PMT AP. From this fit 
$\tau_3$=1399$\pm$20(PMT syst.)$\pm$8(fit syst.)$\pm$6(TPB syst.)$\pm$7(AP syst.)~ns, where errors are evaluated by performing the fit separately on individual PMTs, varying the fit range, and varying the TPB fluorescence decay time and times of the AP distributions within uncertainties. This result is stable throughout the analyzed dataset (Fig.~S1 in Supplemental Material~[]), and consistent with the literature value of 1300$\pm$60~ns~\cite{heindl}.

\begin{figure}[htb]
    \includegraphics[width=8.6cm]{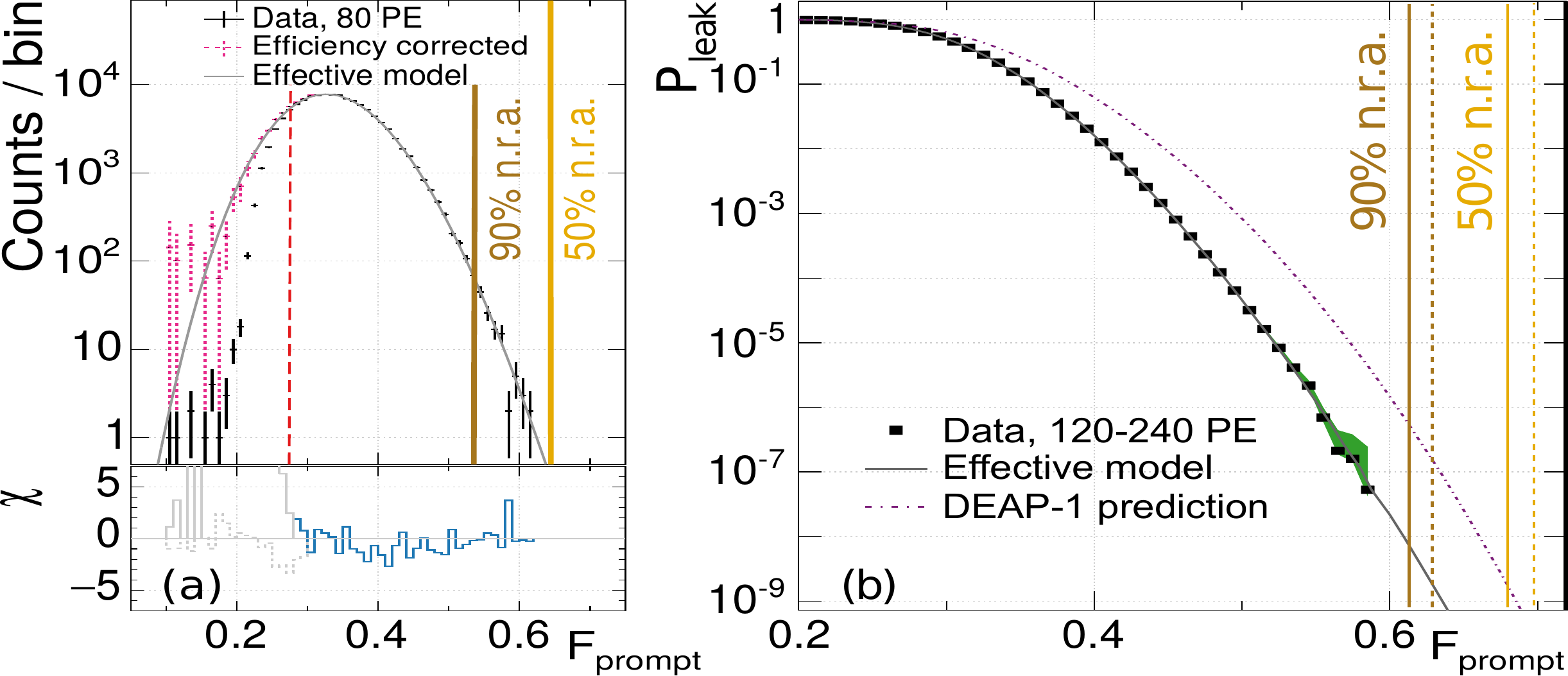}
    \caption{
(a) \fprompt\ vs.~PE distribution slice at 80~PE, with and without the trigger efficiency correction, is shown together with the effective model fit (performed above the red dashed line, indicating the \fprompt\ value below which the trigger efficiency is $<$100\%).
The brown and orange lines correspond to 90\% and 50\% NRA. 
(b) Data and model for the 120--240~PE range with 1.87972$\times$10$^7$ events, represented as leakage probability above given \fprompt. A conservative projection from DEAP-1~\cite{psd} is also shown with its NRA lines (dashed).}
    \label{fig:psd}
\end{figure}

\nuc{Ar}{39} \textbeta\ decay, the dominant source of scintillation events, results in low-\fprompt\ ERs.
 In order to define an \fprompt\ cut constraining the leakage of \nuc{Ar}{39} events into the NR band, the \fprompt\ distribution of ERs and its energy dependence were fitted with an 11-parameter empirical model of \fprompt\ vs.~PE, based on a widened Gamma distribution,
$\textrm{PSD}(n, f) = \Gamma(f; \overline{f}(n), b(n) ) \otimes \textrm{Gauss}(f; \sigma(n))$,
where $b(n)=a_0 + \frac{a_1}{n} + \frac{a_2}{n^2}$, $\sigma(n)=a_3 + \frac{a_4}{n} + \frac{a_5}{n^2}$ and $\overline{f}(n)$ is parametrized as $a_6 + \frac{a_7}{n-a_8} + \frac{a_9}{(n-a_{10})^2}$. The 2-dimensional fit of the model to the data (80--260~PE) has $\chi^2_{ndf}$ of 5581/(5236-11). 
Each PE bin contributes approximately equally to $\chi^2$; as an example, a 1-dimensional slice at 80~PE is shown in Fig.~\ref{fig:psd}(a).
The PSD leakage measured in the 120--240~PE window with a 90\% NR acceptance (NRA) is shown in Fig.~\ref{fig:psd}(b). The extrapolated leakage is approximately 10$\times$ lower than projected in the DEAP-3600 design~\cite{psd}.
As further PSD leakage reduction is expected from SPE counting~\cite{singlepe}, the original goal of a 120~PE analysis threshold in 3~tonne-years will likely be surpassed.

The energy calibration uses internal backgrounds and external radioactive sources.
The internal calibration uses \textbeta's from \nuc{Ar}{39} decay, with an endpoint of 565~keV and uniformly distributed in the detector (as WIMP-induced NRs would be). The external calibration uses a \nuc{Na}{22} source, which produces 1.27~MeV \textgamma's and a 30--50~keV photoabsorption feature near the AV surface.
The simulated spectra of \nuc{Ar}{39} and \nuc{Na}{22} are fit to the data (separately, because of different spatial distributions) to find the energy response function relating T$_{\mathrm{eff}}$~[\kevee] (electron-equivalent energy) to detected PE,
\begin{equation}\label{eq:eresp}
N_{\mathrm{PE}}(T_{\mathrm{eff}}) = c_0 + c_1 T_{\mathrm{eff}} + c_2 T_{\mathrm{eff}}^2,
\end{equation}
where $c_0$=1.2$\pm$0.2~PE, $c_1$=7.68$\pm$0.16~PE \kevee$^{-1}$ and $c_2$=-(0.51$\pm$2.0)$\times10^{-3}$~PE \kevee$^{-2}$.
The offset $c_0$ is fixed to values returned by analysis of mean pretrigger window charge for each run. The \nuc{Ar}{39} fit result constitutes the nominal calibration, while the \nuc{Ar}{39}--\nuc{Na}{22} fit parameter differences, determined from a pair of runs taken just after the 2nd fill, are combined with the statistical uncertainties and used as systematic uncertainties from position and model dependence on $c_{1,2}$.

\begin{figure}[htb]
    \centering
    \includegraphics[width=8.6cm]{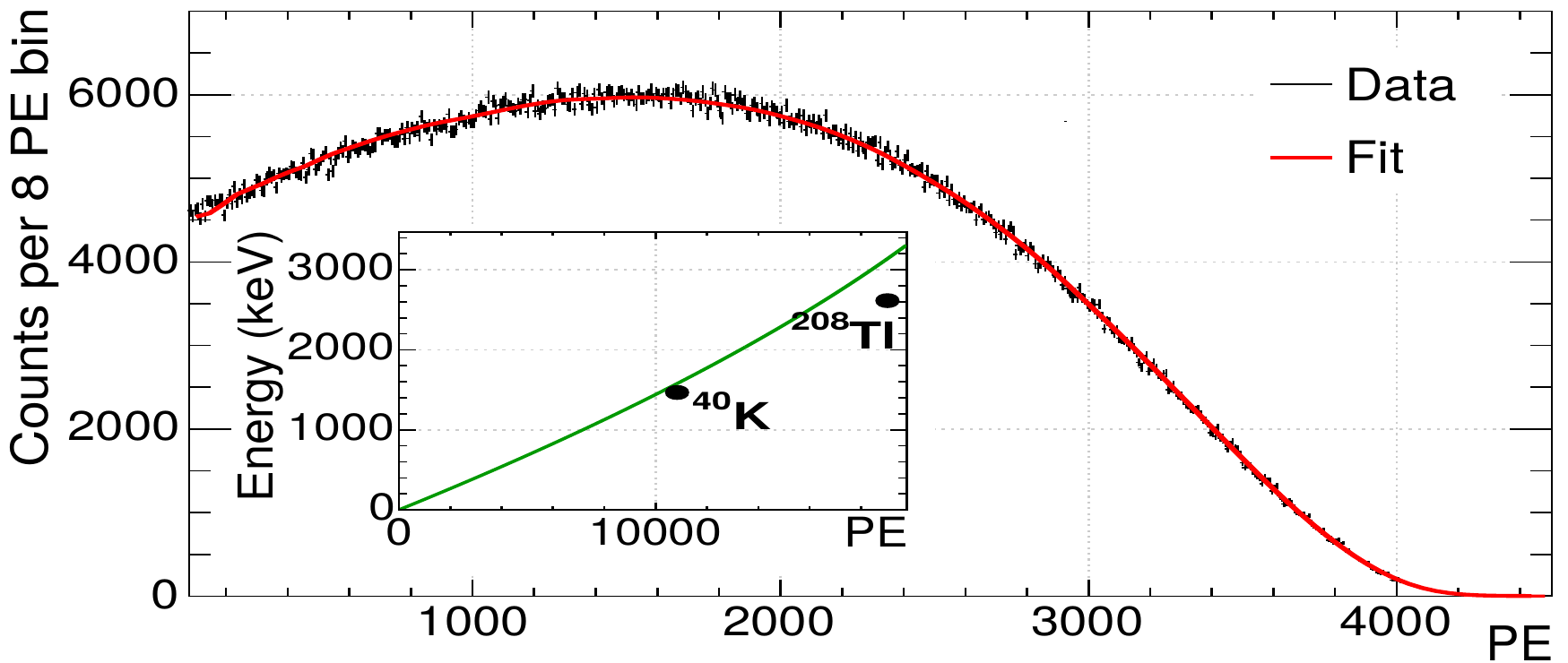}
    \caption{Measured, trigger-efficiency-corrected \nuc{Ar}{39} \textbeta\ spectrum from a subset of data and the fit function (red) based on simulation, with $\chi^2_{ndf}=1.02$. The inset shows the energy response function, Eq.~(\ref{eq:eresp}), from the \nuc{Ar}{39} fit, and, as a cross-check, \textgamma\ lines from \nuc{K}{40} and \nuc{Tl}{208}. \nuc{Tl}{208} diverges from the function because of PMT and DAQ non-linearity.}
\label{fig:ene}
\end{figure}

The final response function is shown in Fig.~\ref{fig:ene} together with the \nuc{Ar}{39} data spanning from below to above the analysis energy window (see Fig.~S2 in Supplemental Material~[] for the \nuc{Na}{22} fit). The energy response function linear terms, $c_1$, for \nuc{Ar}{39} and \nuc{Na}{22} agree within errors. The response function is extrapolated to compare with high-energy \textgamma\ lines, see Fig.~\ref{fig:ene}.

The light yield (LY) at 80~PE is 7.80$\pm$0.21(fit syst.)$\pm$0.22(SPE syst.)~PE/\kevee, where the latter uncertainty is from SPE calibration.

A Gaussian resolution function is used in the fit, with $\sigma^2 = c_0 + p_1 (\mathrm{PE}-c_0)$.
The resolution at 80~PE extrapolated from best fit values for \nuc{Ar}{39} and \nuc{Na}{22} is 20$\pm$1\% and 21$\pm$1\%, respectively.
A lower bound on the energy resolution at 80~PE is 12\% ($p_1=1.185$), determined from counting statistics widened by the measured in-situ SPE charge resolution. 
Due to the steeply falling WIMP-induced spectrum, broader resolutions imply stronger limits at low WIMP masses. Thus using this lower bound is conservative.

NRA of the \fprompt\ cut is determined from a simulation of $^{40}$Ar recoils distributed uniformly in LAr.
The simulation assumes the quenching factor (QF, the LY of NRs relative to ERs) measured by SCENE~\cite{scene} at zero electric field, and the $\frac{I_1}{I_3}$ energy dependence required to reproduce the reported median {\tt f$_{90}$} values; SCENE uncertainties are propagated through the analysis. The simulation applies the full response of the detection and analysis chain, including all noise components affecting the \fprompt\ distribution shape and width.
PMT AP is the dominant effect contributing to shifting \fprompt\ relative to the intrinsic value~\cite{ap}, with an average AP probability of (7.6$\pm$1.9)\%~\cite{det}, $\sim$5$\times$larger than in SCENE. This 7.6\% produces a proportional 5\% shift in the median \fprompt.
Comparison of external neutron AmBe source data with a simplified detector simulation shows qualitative agreement
 and serves as a validation (Fig.~S3 in Supplemental Material~[]). AmBe data is not used directly to model the WIMP-induced NRA as 59\% of AmBe events in the 120--240~PE window contain multiple elastic neutron scatters.
 \begin{figure}[htb]
    \centering
    \includegraphics[width=8.6cm]{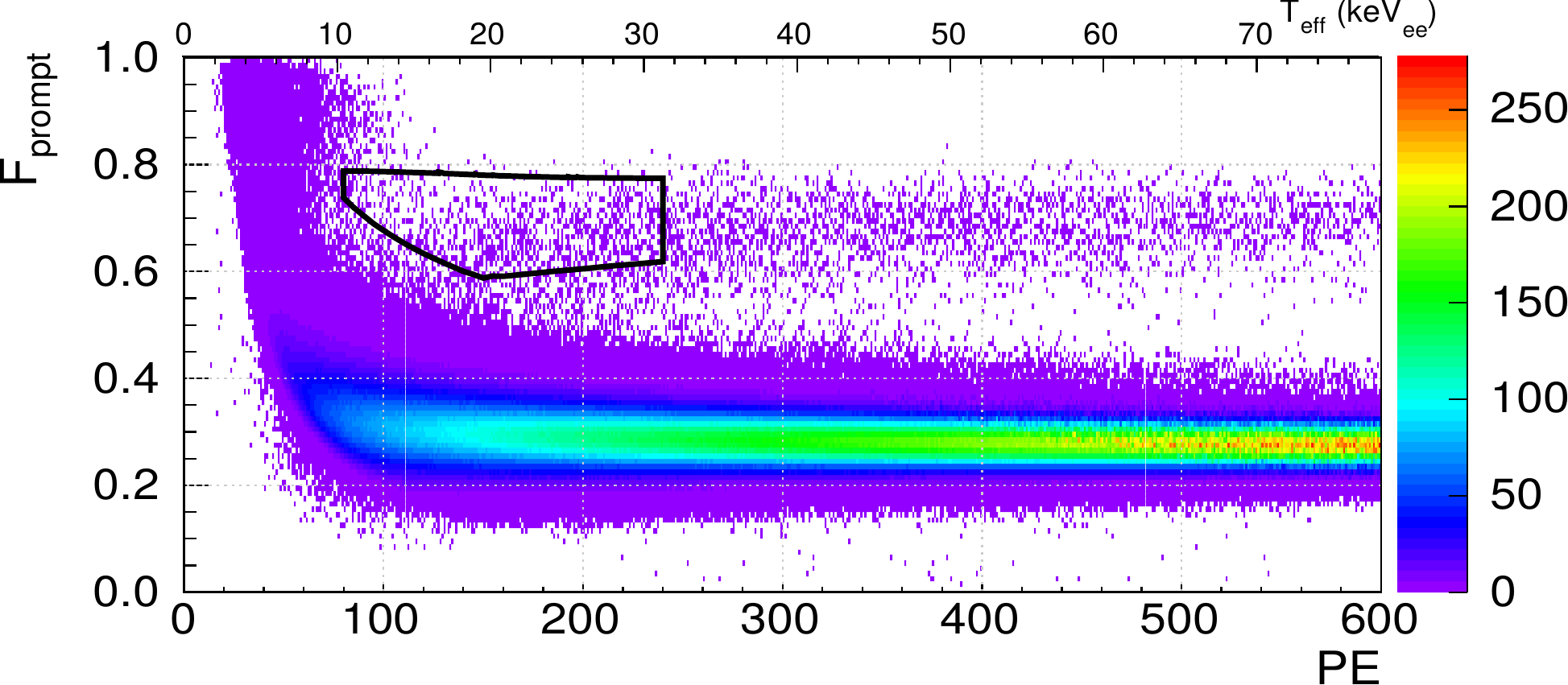}
    \caption{AmBe source data after cuts, with the WIMP search ROI (black box).
} \label{fig:nrambea}
\end{figure}

The region-of-interest (ROI), see Fig.~\ref{fig:nrambea}, was defined by allowing for an expectation of 0.2 leakage events from the \nuc{Ar}{39} band, determined with the PSD model.
The smaller number of \nuc{Ar}{39} events in the short exposure and the low \fprompt\ leakage allowed us to set the threshold at 80~PE (10~\kevee), lower than the nominal 120~PE originally projected~\cite{psd}.
Above 150~PE the lower limit on \fprompt\ is chosen to remove 5\% of NRs in each bin.
The ROI also has a maximum \fprompt\ chosen to remove 1\% of NRs in each bin. 
The maximum energy of 240~PE, where the nominal design value was used (subject to future optimization), reduces possible backgrounds from the surface \textalpha\ activity~\cite{deap1Rn}.

The first LAr fill took approximately 100~days between May and mid-August 2016.
For the majority of this time, Ar gas was introduced into the detector from the purification system for cooling.
In the final phase of the fill, shortly following the discussed dataset, a leak in the detector neck contaminated LAr with clean Rn-scrubbed N$_2$. The detector was subsequently emptied and refilled, and it has been taking data since Nov.~1, 2016, with a slightly lower liquid level.

Here we focus on Aug.~5--15 (9.09~days), when the detector contained a constant LAr mass.
A sharp drop in rate between PMTs facing the liquid vs. the vapour space, permits determination of the fill level, 590$\pm$50~mm above the AV centre, and the full LAr mass: 3322$\pm$110~kg (Fig.~S1 in Supplemental Material~[]).

Calibrations were performed after the 2nd fill: 23~h of \nuc{Na}{22} (Nov.~3-4) and 65~h of AmBe data (Dec.~2-4).

Data were analyzed from runs where
(1) the difference between the maximum and minimum AV pressures
corresponded to $<$10~mm change in the liquid level
and (2) there were no intermittently misbehaving PMTs, i.e. no PMT read $<$50\% of its average charge, determined from approximately 5~minute samples. Independently, during this dataset one PMT was turned off (and has since returned to operation).
In all cases, pressure excursions were correlated with periods of the cryocoolers operating at reduced power.
Out of 8.55~d of physics runs, 2.92~d are removed by failing both criteria and 0.91~d by failing criterion 2 alone.
The remaining 4.72~d contained a total deadtime of 0.28~d, due to 17.5~\textmu s deadtime after each trigger, resulting in a 4.44~d livetime.

\begin{figure}[htb]
    \centering
    \includegraphics[width=8.6cm]{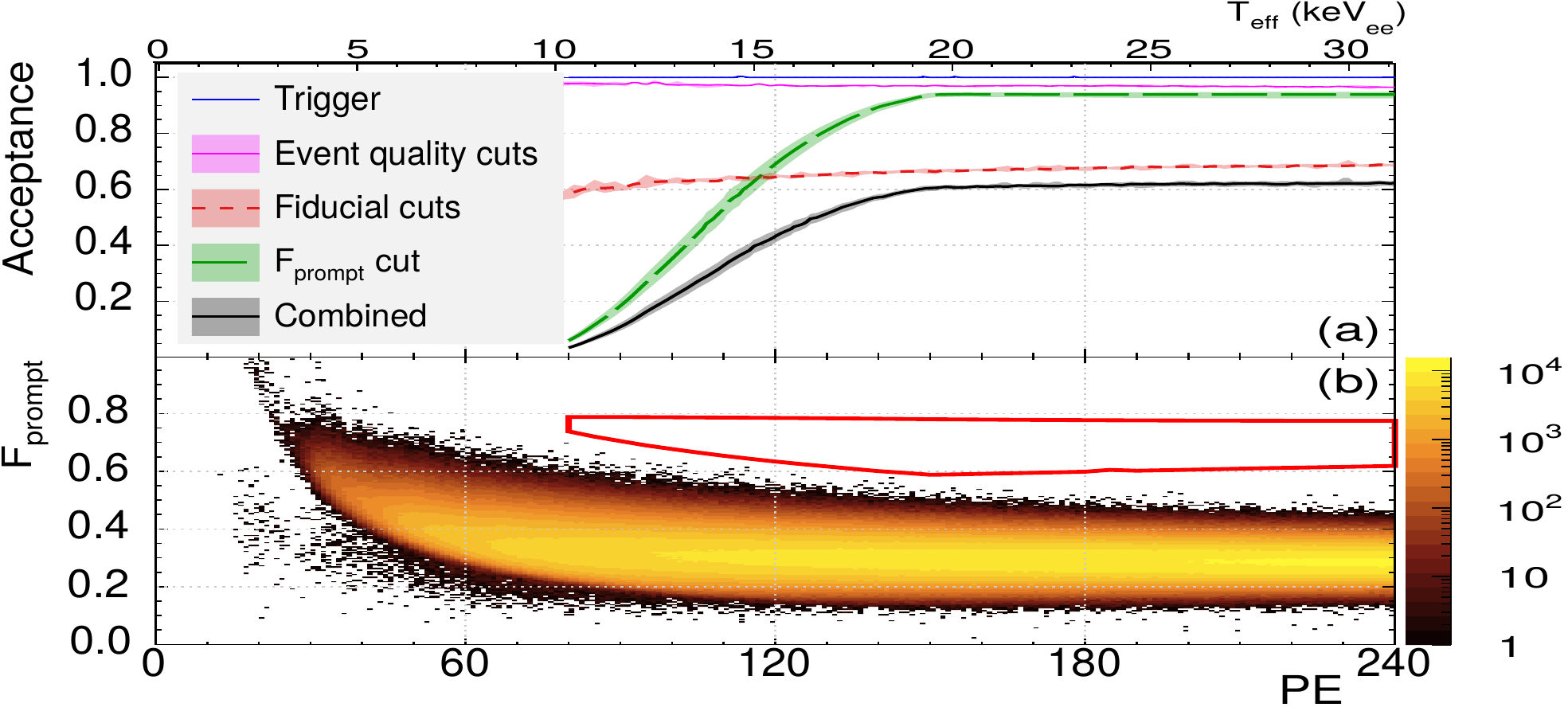}
    \caption{(a) The acceptance in the 80--240~PE window, with systematic errors (maximum variation about the weighted mean, run-by-run). Uncertainties on trigger acceptance and \fprompt\ cut acceptance are discussed in the text. (b) \fprompt\ vs.~PE for events passing cuts, with the WIMP search ROI (red).}
    \label{fig:dataplusroi}
\end{figure}

Acceptance for WIMP-induced NR events (Fig.~\ref{fig:dataplusroi}(a)) is determined using a combination of (uniformly distributed) \nuc{Ar}{39} events and simulation of \fprompt\ for NRs. The sample of \nuc{Ar}{39} single-recoils is obtained first by applying low-level cuts to remove events (1) from DAQ calibration, (2) from pile-up or (3) highly asymmetric ($>$40\% of charge in a single PMT) e.g. Cherenkov events in LGs and PMTs. The approach of measuring acceptance for NRs using ERs is used since none of the cut variables depend on the pulse time information, only \fprompt\ does, which is handled separately. The \fprompt\ simulation for NRs is validated by comparison with the AmBe data. See Table~S1 in Supplemental Material~[] for the impact breakdown of run selection and cuts.

Quality cuts are applied to $^{39}$Ar events within the energy window in order to determine the ER acceptance:
the event time cut requires the scintillation peak positioned early in the waveform (for reliable \fprompt\ evaluation), cuts on the fraction of charge in the brightest PMT and on the neck veto remove high-charge AP triggering the detector as well as light emission in the AV neck (e.g. Cherenkov). We have identified a class of background events originating in the neck region and are characterizing it for future larger-exposure searches. 

The fiducial acceptance is determined relative to the events remaining after the quality cuts.
Fiducialization employs low-level PE ratio cuts. These are that the fraction of scintillation-induced (AP corrected) PE~\cite{ap,singlepe} in the PMT which detects the most light be $<$7\%, and that the fraction of charge in the top 2 PMT rows be $<$5\%. These variables are strongly correlated with the radial and vertical event positions, respectively, and so reject events at the surface of the detector and in the neck.  The volume, after cuts on these variables (Table~S1 in Supplemental Material~[]), corresponds roughly to a sphere of radius $\sim$773~mm, truncated at the LAr level (z$\approx$590~mm). The fiducial mass, 2223$\pm$74~kg, is determined from the full LAr mass and measured acceptance of the fiducialization cuts. The expected  \nuc{Ar}{39} activity contained therein is 2245$\pm$198~Bq~\cite{ar39}, consistent with the fiducial rate observed, 2239$\pm$8~Hz.

Position reconstruction algorithms in this analysis were used only as a cross-check (Fig.~S4 in Supplemental Material~[]).

The main background sources are \textalpha\ activity, neutrons, and leakage from \nuc{Ar}{39} and other ERs. As external backgrounds contributions to this early analysis are negligible, we have not yet determined their distributions.

\nuc{Rn}{222}, \nuc{Po}{218} and \nuc{Po}{214} \textalpha\ decays are identified in the LAr bulk as high-energy peaks or based on delayed coincidences, \textalpha-\textalpha\ (\nuc{Rn}{222}-\nuc{Po}{218} and \nuc{Rn}{220}-\nuc{Po}{216}) or \textbeta-\textalpha\ (\nuc{Bi}{214}-\nuc{Po}{214}), resulting in activities: (1.8$\pm$0.2)$\times$10$^{-1}$~\textmu Bq/kg of \nuc{Rn}{222}, (2.0$\pm$0.2)$\times$10$^{-1}$~\textmu Bq/kg of \nuc{Po}{214}, and (2.6$\pm$1.5)$\times$10$^{-3}$~\textmu Bq/kg of \nuc{Rn}{220} (Fig.~S5 in Supplemental Material~[]). For comparison approximate values from other experiments are:  66~\textmu Hz/kg of \nuc{Rn}{222} and 10~\textmu Hz/kg of \nuc{Rn}{220} in LUX~\cite{luxrn}, 6.57~\textmu Bq/kg of \nuc{Rn}{222} and 0.41~\textmu Bq/kg of \nuc{Rn}{220} in PandaX-II~\cite{pandaxrn}, and 10~\textmu Bq/kg of \nuc{Rn}{222} in XENON1T~\cite{xenon1trn}.
The out-of-equilibrium \nuc{Po}{210} activity is determined with a fit of simulated spectra to the data: 0.22$\pm$0.04~mBq/m$^2$ on the AV surface and $<$3.3~mBq in the AV bulk (Fig.~S6 in Supplemental Material~[]).

(\textalpha, n) reactions and spontaneous fission in the PMTs is the expected dominant source of neutron events. It is constrained with measurements of the 2614~keV and 1764~keV \textgamma-rays from the \nuc{Th}{232} and \nuc{U}{238} decay chains, respectively.
In-situ activities of both decay chains agree within a factor of two with a simulation based on the screening
results.
Neutron backgrounds are also constrained by searching for NRs followed by capture \textgamma's, with efficiency calibrated using neutrons from an AmBe source deployed near the PMTs. No neutron candidates were seen in 4.44~d (80--10000~PE, no fiducial cuts), which is consistent with the assay-based expectation.

Systematic uncertainties in the WIMP cross section limit include uncertainties in the NR energy response, total exposure, and cut acceptance (see Fig.~\ref{fig:dataplusroi}(a)).
The \fprompt\ cut acceptance uncertainty is determined from uncertainties in the simulation parameters, including $\frac{I_1}{I_3}$ (derived from the SCENE {\tt f$_{90}$} measurements~\cite{scene}), $\tau_3$ ($\pm$70\,ns, from the difference between SCENE and this work), and the AP probability.
The main uncertainty is from the NR energy response.
This is dominated by uncertainties in Eq.~(\ref{eq:eresp}), followed by uncertainties in the NR QF. SCENE reports two energy-dependent NR QFs which differ due to non-unitary recombination at null field: $\mathcal{L}_{\mathrm{eff},^{83m}\mathrm{Kr}}$ (the NR LY relative to that from a \nuc{Kr}{83m} ER calibration) and $\mathcal{L}$ (the Lindhard-Birks QF describing the suppression of photon and ionized electron production)
We varied the Lindhard-Birks QF fit to $\mathcal{L}$ to account for the uncertainty of normalizing NR LY relative to the \nuc{Ar}{39} spectrum rather than to \nuc{Kr}{83m} calibration, as SCENE did, using the NEST model~\cite{nest}, fitting Thomas-Imel and Doke-Birks recombination parameters to SCENE's $\mathcal{L}_{\mathrm{eff},^{83m}\mathrm{Kr}}$ values.
These factors, along with the uncertainty in Birks' constant reported by SCENE and the difference between $\mathcal{L}$ and $\mathcal{L}_{\mathrm{eff},^{83m}\mathrm{Kr}}$ were included in the overall QF uncertainty.

\begin{figure}[htb]
    \centering
    \includegraphics[width=8.6cm]{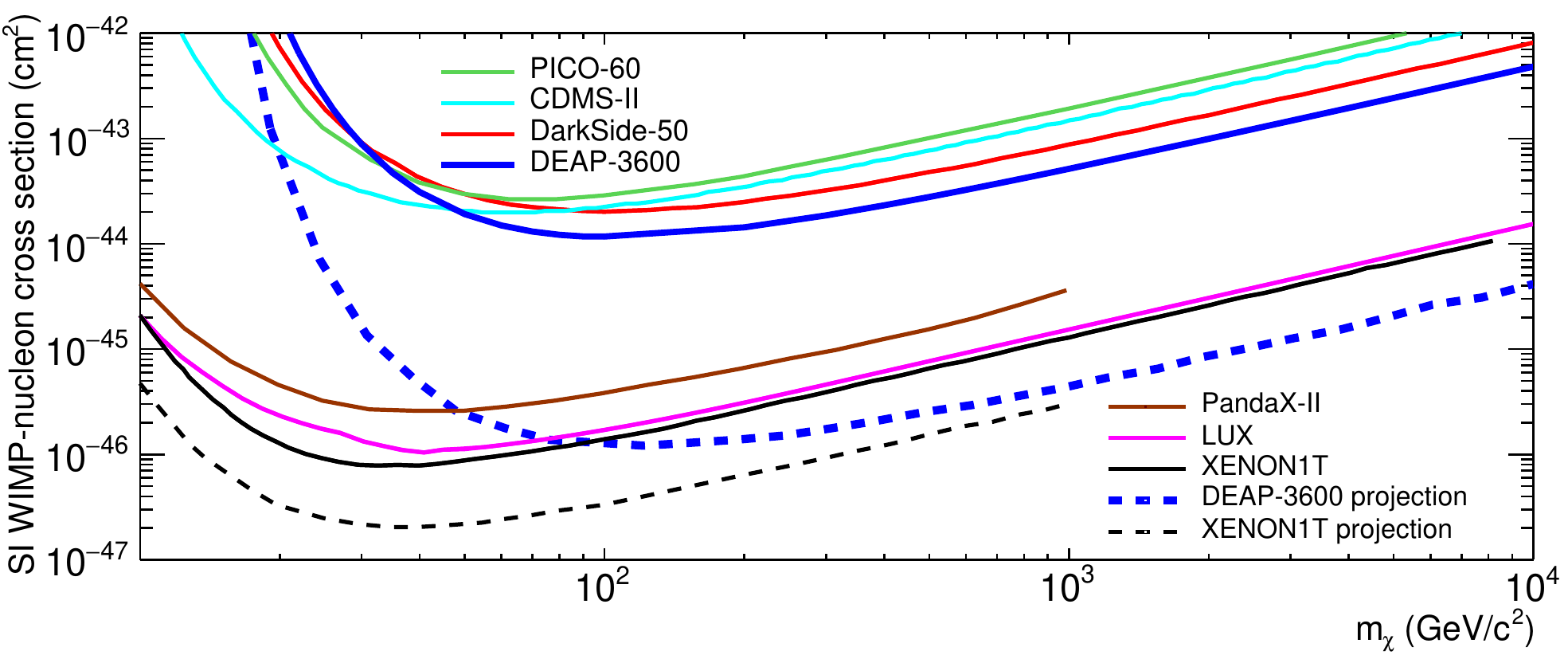}
    \caption{Spin-independent WIMP-nucleon cross section 90\%~C.L. exclusion from 4.44~live days of DEAP-3600 data. Also shown are current results from other searches~\cite{xenon1t,lux,pandax,ds50,cdms,pico}, and projections for XENON1T and DEAP-3600 (a 3~tonne-year background-free exposure with a 15~\kevee\ threshold).}
    \label{fig:exclusion}
\end{figure}

No events are observed in the ROI, see Fig.~\ref{fig:dataplusroi}(b). Figure~\ref{fig:exclusion} shows the resulting limit on the spin-independent WIMP-nucleon scattering cross section, based on the standard DM halo model~\cite{mccabe}. The 90\%~C.L. upper limit is derived employing the Highland-Cousins method~\cite{hc}.
For a more conservative limit, the predicted \nuc{Ar}{39} leakage was not subtracted. This analysis was not blind.

DEAP-3600 achieved 7.8~PE/\kevee\ LY at the end of the detector fill without recirculation, and demonstrated better-than-expected PSD (permitting a 37~\kevr\ threshold), with promising \textalpha\ and neutron background levels. Analysis of the first 4.44~d of data results in the best limit at low energies on discrimination of \textbeta-decay backgrounds using PSD in LAr at 90\% NRA, with measured leakage probability of $<$1.2$\times$10$^{-7}$ (90\%~C.L.) in the energy window 15--31~\kevee\ (52--105~\kevr).  This measurement has lower threshold than DEAP-1~\cite{psd} and higher statistics than DarkSide-50~\cite{ds50}. 
After NR selection cuts no events are observed, resulting in the best spin-independent WIMP-nucleon cross section limit measured in LAr of $<$1.2$\times$10$^{-44}$~cm$^2$ for a 100~GeV/c$^2$ WIMP (90\%~C.L.)\footnote[3]{After submission of this paper, DarkSide-50 announced new results~\cite{ds50new}. DEAP-3600 remains the most sensitive non-Xe search in the 48--90~GeV/c$^2$ mass range.}.  
Data collection has been ongoing since Nov.~2016 and forms the basis for a more sensitive DM search currently in progress.

\begin{acknowledgments}
This work is supported by the Natural Sciences and Engineering Research Council of Canada, the Canadian Foundation for Innovation (CFI), the Ontario Ministry of Research and Innovation (MRI), and Alberta Advanced Education and Technology (ASRIP), Queen's University, University of Alberta, Carleton University, DGAPA-UNAM (PAPIIT No. IA100316 and IA100118) and Consejo Nacional de Ciencia y Tecnolog\'ia (CONACyT, Mexico, Grant No. 252167), European Research Council (ERC StG 279980), the UK Science \& Technology Facilities Council (STFC) (ST/K002570/1), the Leverhulme Trust (ECF-20130496). Studentship support by the Rutherford Appleton Laboratory Particle Physics Division, STFC and SEPNet PhD is acknowledged.
We thank SNOLAB and its staff for support through underground space, logistical and technical services. SNOLAB operations are supported by CFI and the Province of Ontario MRI, with underground access provided by Vale at the Creighton mine site.
We thank Compute Canada, Calcul Qu\'ebec and the Center for Advanced Computing at Queen's University for providing the excellent computing resources required to undertake this work.
\end{acknowledgments}
\bibliographystyle{apsrev4-1}





\widetext
\clearpage
\begin{center}
\textbf{\large Supplemental Materials: First results from the DEAP-3600 dark matter search with argon at SNOLAB}
\end{center}

\setcounter{equation}{0}
\setcounter{figure}{0}
\setcounter{table}{0}
\setcounter{page}{1}
\makeatletter
\renewcommand{\theequation}{S\arabic{equation}}
\renewcommand{\thefigure}{S\arabic{figure}}
\renewcommand{\thetable}{S\arabic{table}}
\renewcommand{\bibnumfmt}[1]{[S#1]}
\renewcommand{\citenumfont}[1]{S#1}
\begin{figure}[htb]
    \centering
    \includegraphics[width=8.6cm]{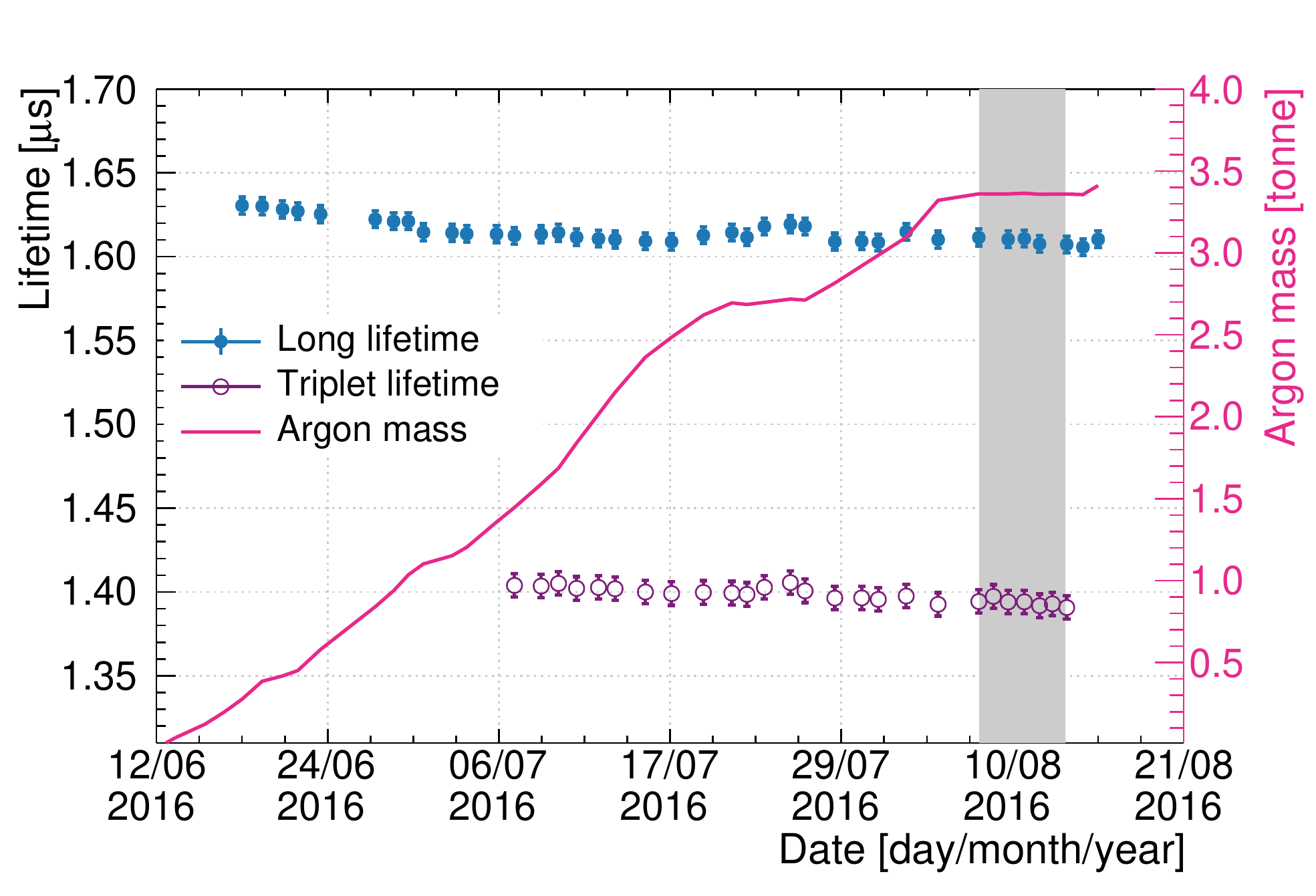}
    \caption{Stability of late light timing measured during the detector fill. \\\\
      The `long lifetime' is determined with a simple `exponential + linear' fit to the summed waveforms from 500~ns to 3000~ns. Such a fit overestimates the triplet lifetime by including effects such as PMT AP and TPB fluorescence in addition to the LAr triplet lifetime. Shown as well is the `triplet lifetime' extracted from the same pulse shapes with a more elaborate fit that accounts for these effects. At low LAr mass, the averaged pulse shapes include a contribution from gaseous Ar scintillation, which also leads to an increase in the fit lifetimes. The triplet lifetime measurements are shown for a period when a sizable amount of LAr was in the detector. The error bars shown are primarily systematic, as the statistical uncertainties from the fits are smaller than the marker size. \\\\
       The grey shaded area represents the dataset used for the dark matter search presented here. The fit time constant is stable within that period to $<$1\%.\\\\
      Argon was delivered as cryogenic liquid, stored underground, purified to sub-ppb impurity levels, scrubbed of Rn~[6] and liquified in the AV. LAr was not recirculated or repurified during the fill.\\\\
      The liquid level and, ultimately, the mass of LAr is inferred from the PMT rates. The refractive index difference between liquid and gaseous Ar is such that scintillation light in the LAr reaching the surface of the liquid with angle of incidence $>$53$^\circ$ is totally internally reflected. This produces rates in the PMTs facing the gas which are 20\% lower than rates of PMTs facing the liquid.}
    \label{fig:triplet}
\end{figure}

\begin{figure}[htb]
    \centering
    \stackinset{l}{0.141\columnwidth}{t}{0.155\columnwidth}{\includegraphics[clip=true, trim=0 0 0 0,width=0.15\columnwidth]{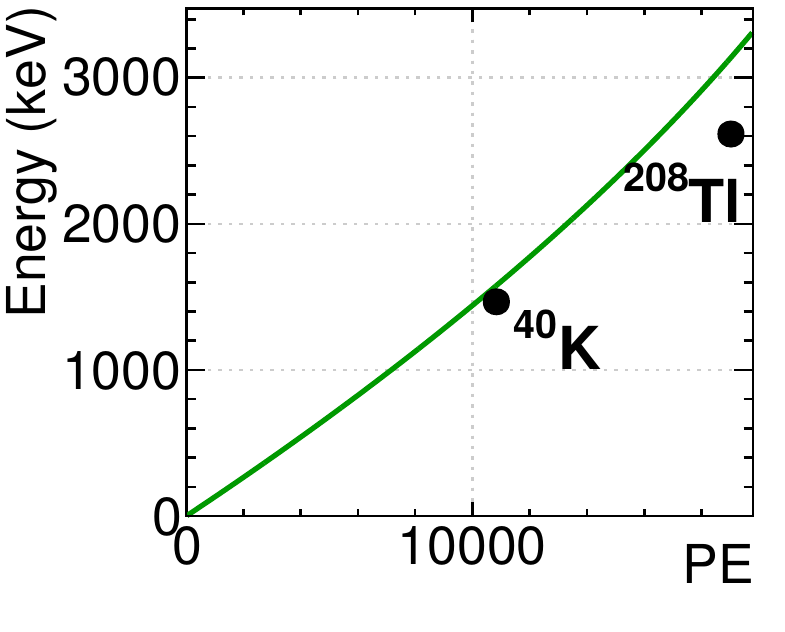}}{
    \includegraphics[width=0.5\columnwidth]{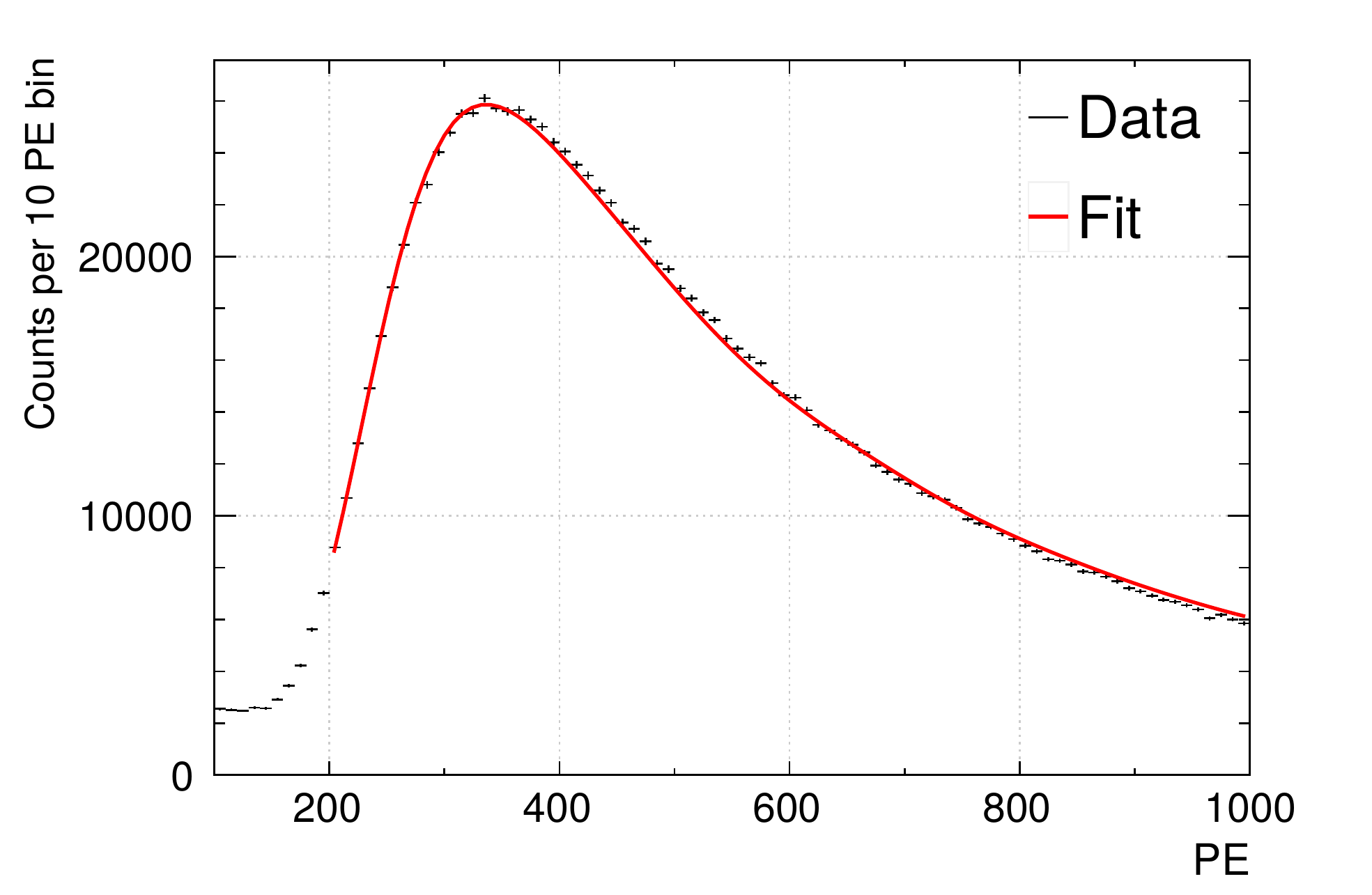}}\includegraphics[width=0.5\columnwidth]{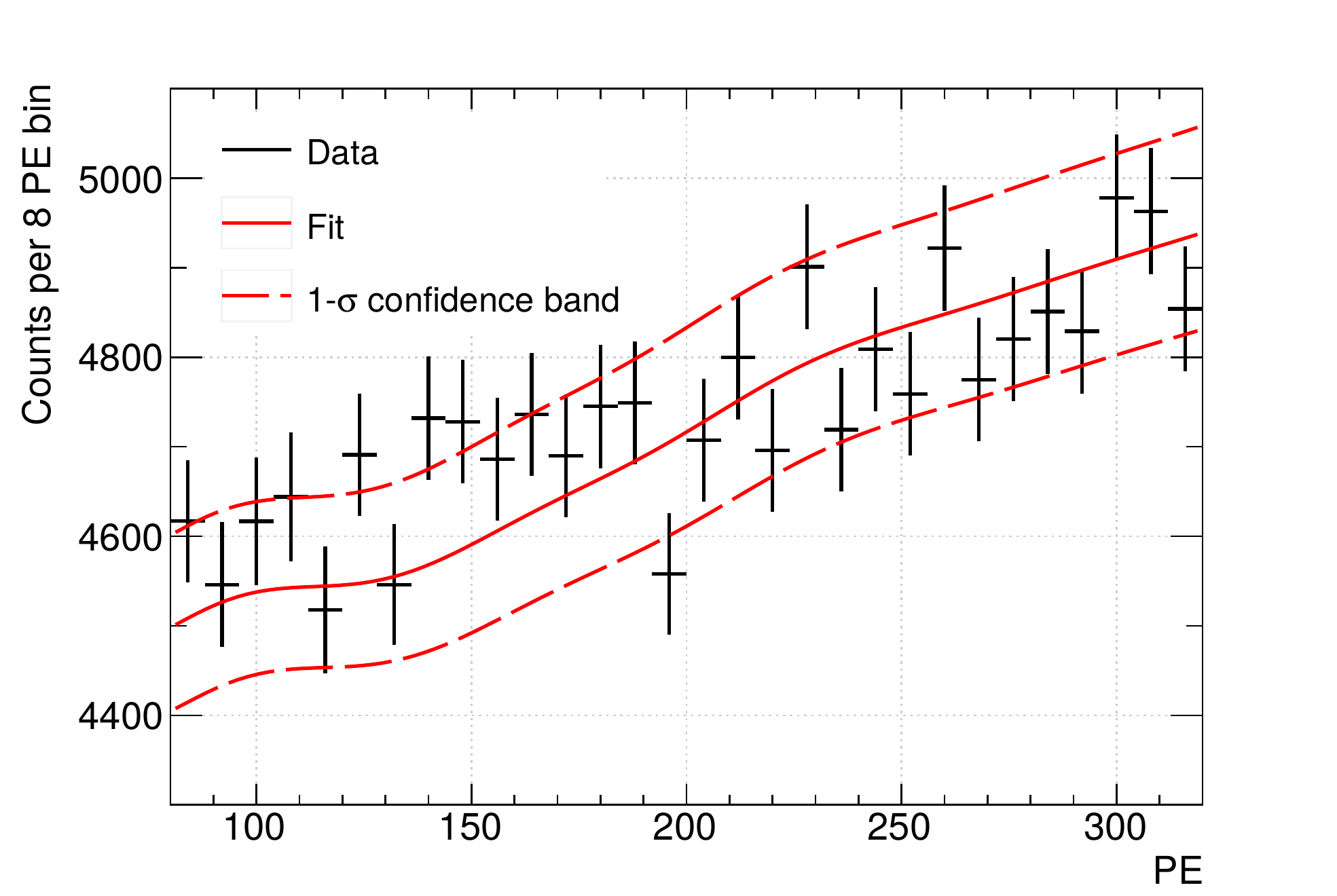}
    \caption{(a) Spectrum collected after the 2nd detector fill with an external \nuc{Na}{22} source overlayed with the fit function (red) based on simulation. The inset shows the global energy response function from the \nuc{Ar}{39} fit, $N_{\mathrm{PE}}(T_{\mathrm{eff}})=c_0+c_1 \mathrm{T_{\mathrm{eff}}}+c_2 \mathrm{T_{\mathrm{eff}}}^2$, with $c_0$=1.2$\pm$0.2~PE, $c_1$=7.68$\pm$0.16~PE \kevee$^{-1}$ and $c_2$=-(0.51$\pm$2.0)$\times10^{-3}$~PE \kevee$^{-2}$.  As a cross-check, on the inset \textgamma\ lines from \nuc{K}{40} and \nuc{Tl}{208} are compared with the extrapolated function;  \nuc{Tl}{208} diverges from the function because of PMT and DAQ non-linearity. (b) \nuc{Ar}{39} data and fit from Fig.~2 zoomed in to the low energy window, with 1-$\sigma$ confidence band shown (dashed).}
\label{fig:na22}
\end{figure}
 \begin{figure*}[htb]
    \centering
    \includegraphics[width=8.4cm]{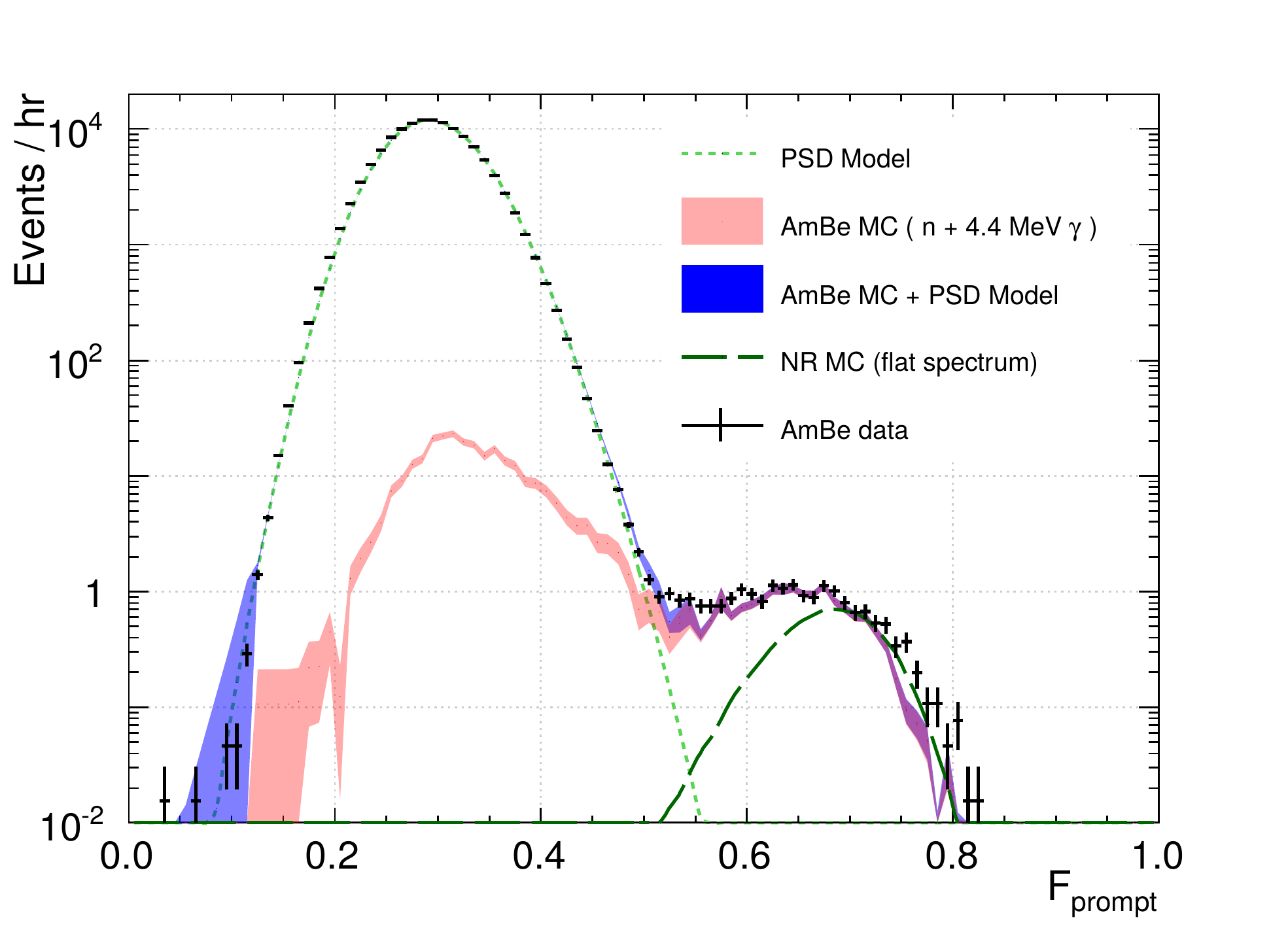}
    \caption{The \fprompt\ distribution for 140$<$PE$<$240 in AmBe calibration data, compared to summed simulated contributions for AmBe neutrons, and 4.4~MeV \textgamma's and the \nuc{Ar}{39} \fprompt\ model normalized to the peak of the distribution. Also plotted is the Monte Carlo simulation of single scatter nuclear recoils, our proxy for WIMP-induced events, with a flat energy spectrum (see legend). Error bars shown on the simulated distributions are statistical, not systematic. \\\\
      The simulation includes neutrons and 4.4~MeV \textgamma's from the AmBe source and, because of CPU time limitations, only considers scattering- or capture-induced \textgamma's for neutrons that entered LAr. This simplifying assumption results in a deficit of events in the simulated data in the gap between both bands due to missing pile-up of neutrons with neutron-induced \textgamma's and, at high \fprompt, due to missing Cherenkov  events (from 4.4~MeV \textgamma's in absence of neutrons in LAr). The peak positions of MC and data distributions are consistent within the systematic uncertainty from the simulation parameters.
} \label{fig:nrambeb}
\end{figure*}

\begin{table}
\footnotesize
\begin{tabular}{clcccc}
\hline   
\hline 
& Cut &  Livetime & \multicolumn{2}{c}{Acceptance \%} & \#ROI events\\
\hline 
\multirow{4}{*}{\rotatebox[origin=c]{90}{\tiny run}}& Physics runs & 8.55 d  &   &  \\ 
& Stable cryocooler & 5.63 d & & \\
& Stable PMT & 4.72 d  &   &  & \\
& Deadtime corrected & 4.44 d & & &  119181 \\
& & & & &  \\
\multirow{3}{*}{\rotatebox[origin=c]{90}{\tiny low level}} & DAQ calibration & & & & 115782 \\
 & Pile-up & & & & 100700 \\
 & Event asymmetry & & && 787 \\
& & & & &  \\
\multirow{4}{*}{\rotatebox[origin=c]{90}{\tiny quality}}& Max charge fraction & & \multirow{2}{*}{99.58$\pm$0.01} & & \multirow{2}{*}{654}\\
& \ per PMT &&&& \\
& Event time & & 99.85$\pm$0.01 & & 652 \\
& Neck veto & & $97.49^{+0.03}_{-0.05}$ & & 23\\
& & & & &  \\
\multirow{4}{*}{\rotatebox[origin=c]{90}{\tiny fiducial}}& Max scintillation PE& & & \multirow{2}{*}{$75.08^{+0.09}_{-0.06}$}  & \multirow{2}{*}{7} \\
& \ fraction per PMT&&&& \\
& Charge fraction in& & & \multirow{2}{*}{$90.92^{+0.11}_{-0.10}$} & \multirow{2}{*}{0} \\
& \ the top 2 PMT rings&&&& \\
& & & & \\
& Total & 4.44 d  & 96.94$\pm$0.03 & $66.91^{+0.20}_{-0.15}$ & 0\\
\hline
\hline
\end{tabular} 
\caption{Run selection criteria and cuts with their effects on livetime, integrated acceptance, the fiducial fraction, and the number of events left in the ROI. The acceptance is calculated individually for each run and then weighted by livetime to provide an overall acceptance with the uncertainties taken as maximum variations about this weighted mean from each run. The fiducial acceptance is determined relative to the events remaining after the quality cuts, which are considered a clean sample of \nuc{Ar}{39} \textbeta's uniformly distributed in LAr. The total number of triggers before any cuts was 1.38$\times$10$^9$, with 6.47$\times$10$^7$ in the 80-240~PE window.}
\label{tab:cuts2}
\end{table}

\begin{figure*}[htb]
    \centering
    \includegraphics[width=0.42\textwidth]{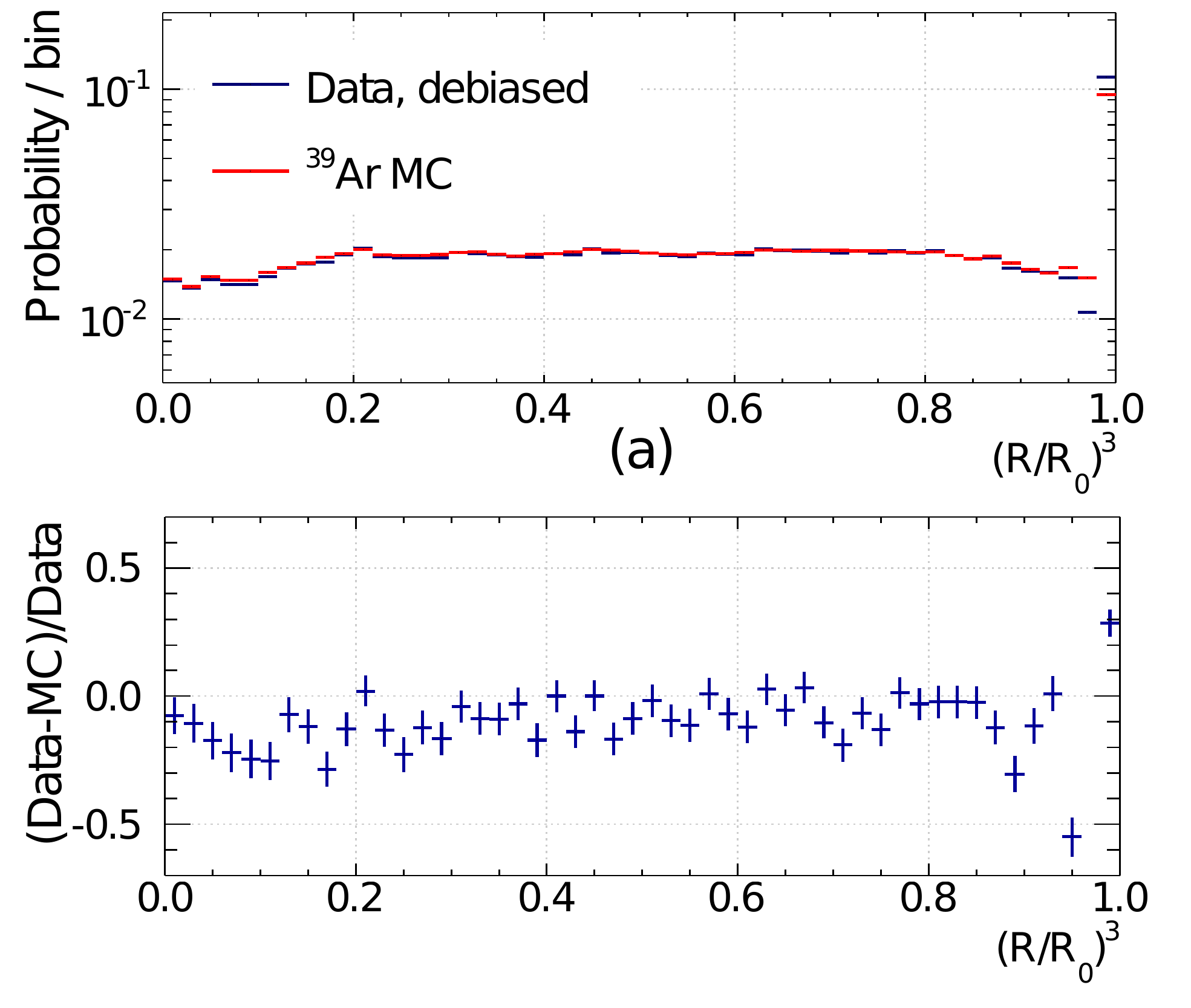}\includegraphics[width=0.42\textwidth]{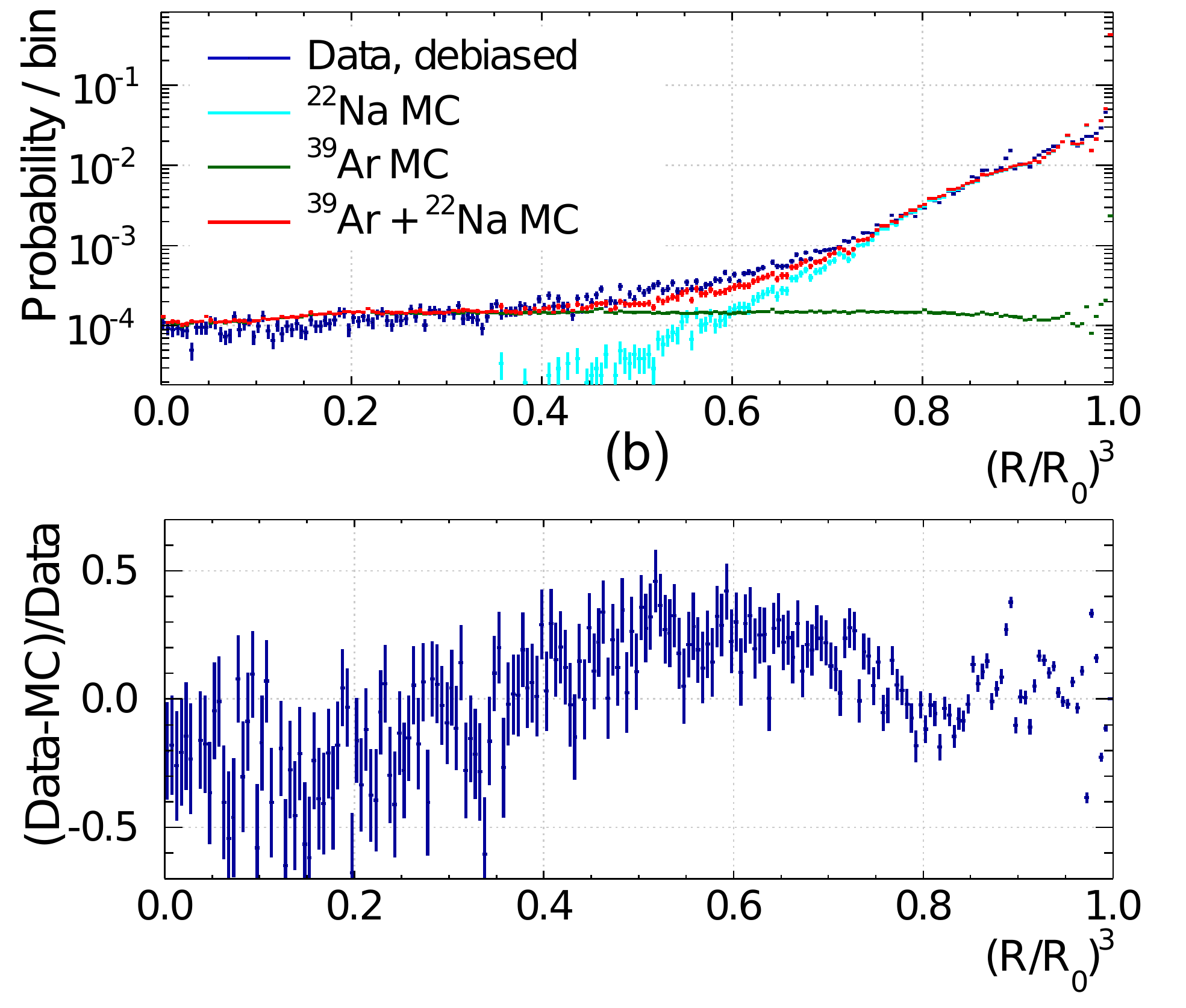}
    \caption{
      A maximum likelihood fitter relies on the full Monte Carlo of the detector, including its optical properties, and minimizes the difference between the observed pattern of PMT charges and the one expected based on a distribution constructed from simulation, under the assumption that the illumination of the detector is symmetric around the axis of the event position vector. Residual position bias is corrected for using the uniformly distributed population of \nuc{Ar}{39} \textbeta's. To study the reconstruction of events from the inner AV surface, as expected for \textalpha\ backgrounds, we apply the \nuc{Ar}{39}-derived calibration to \nuc{Na}{22} events, which are strongly peaked near the surface.\\\\
      Reconstructed radii of (left) \nuc{Ar}{39} uniformly distributed in the detector and (right) tagged events from an external \nuc{Na}{22} calibration source after correcting for radial bias in data (blue), \nuc{Na}{22} Monte Carlo (cyan), distribution of random coincidences of the source tag with \nuc{Ar}{39} decays (green) and the sum of both \nuc{Ar}{39} and \nuc{Na}{22} distributions (red). Residuals are displayed in the bottom row, showing qualitative agreement.\\\\
      Position reconstruction algorithms will be used to further reduce backgrounds in longer exposure runs.
    }
    \label{fig:ar39na22pos}
\end{figure*}

\begin{figure*}[htb]
    \centering
    \includegraphics[width=8.6cm]{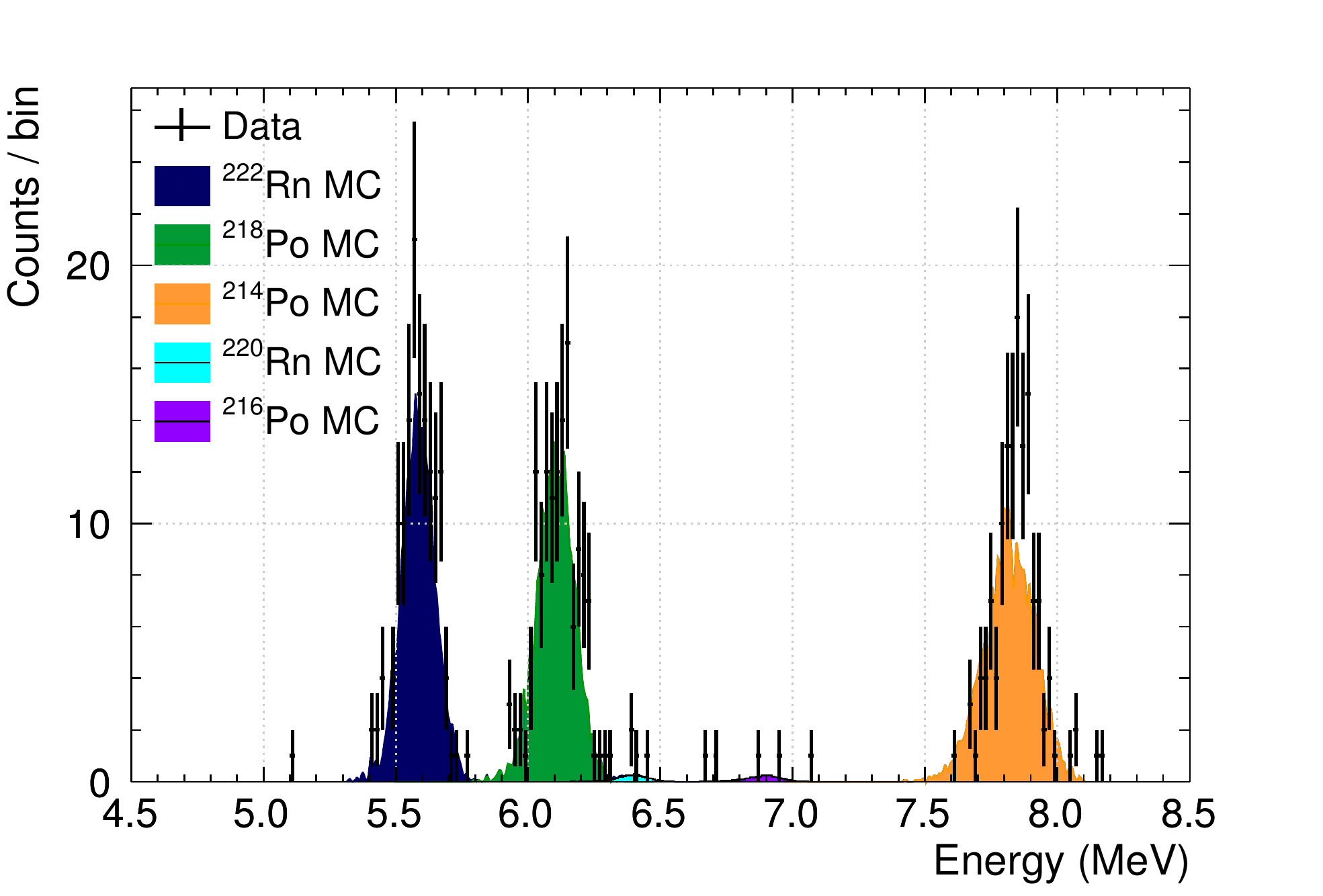}
    \caption{Peaks from tagged alphas in \nuc{Rn}{222} and \nuc{Rn}{220} chains in the detector data overlayed with Monte Carlo distributions. The activity of \nuc{Po}{214} in the bulk is consistent with the earlier part of the chain, indicating that it is mostly mixed within the LAr volume.}
    \label{fig:alpha1}
\end{figure*}

\begin{figure*}[htb]
    \centering
    \includegraphics[width=8.6cm, trim=0cm 0cm 0 0]{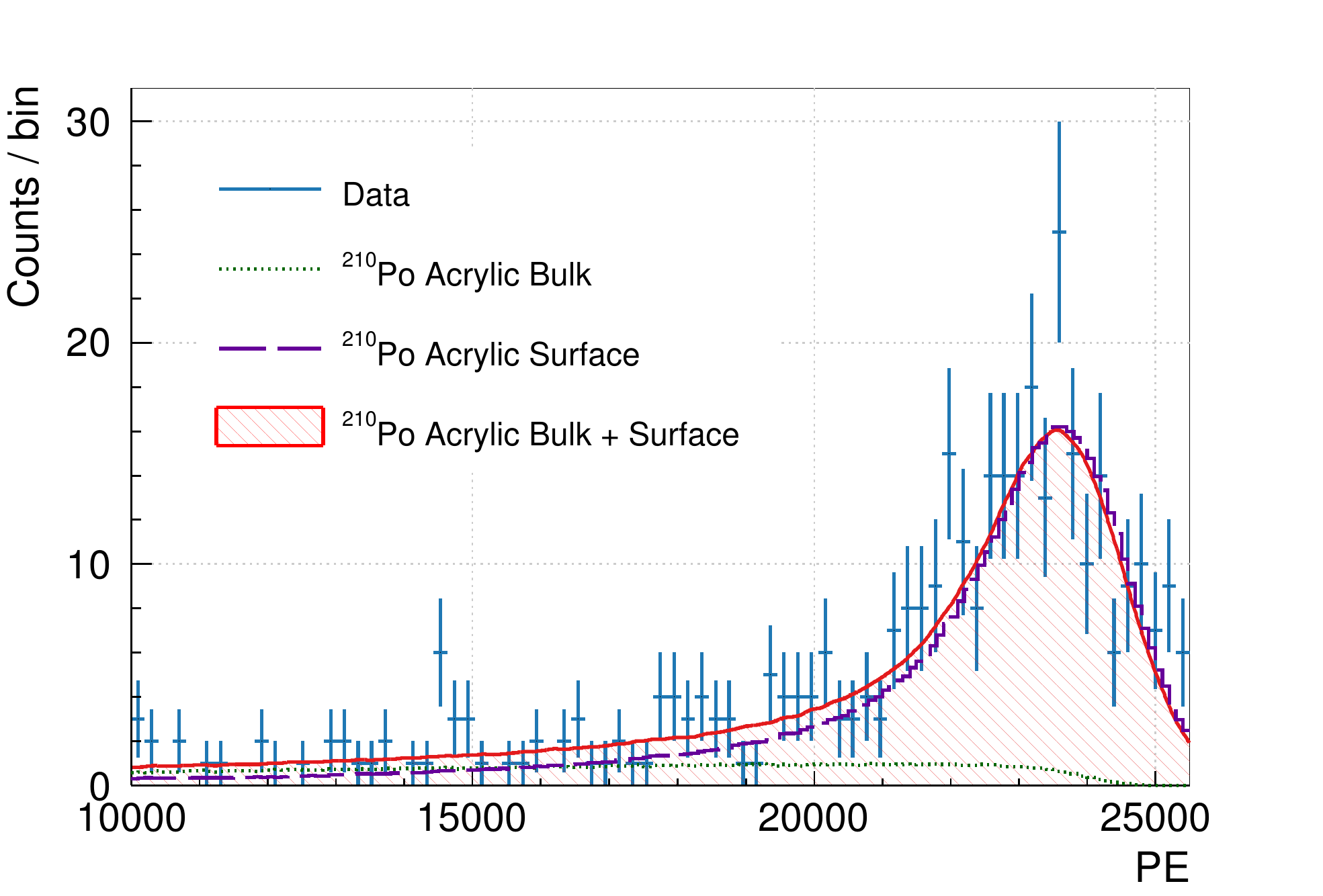}
    \caption{The out-of-equilibrium \nuc{Po}{210} activity is identified by degraded energies characteristic of \textalpha's coming from below the TPB layer. The \nuc{Po}{210} peak in the data overlayed with simulated distributions of contamination present on the acrylic surface (dashed, purple), contamination uniformly distributed in a surface layer of 80~microns depth (dotted, green), and a fit combining both distributions (solid, red). The fit assumes no additional backgrounds in the wide range; therefore the result for the bulk contamination is considered an upper limit.\\\\
      To avoid \nuc{Rn}{222}/\nuc{Pb}{210} contamination of the bulk acrylic and the AV-TPB interface, the inner 0.5~mm layer of the AV was removed
in-situ after construction. The Rn exposure was then strictly limited, with the AV and the access glovebox purged with Rn-scrubbed N$_2$, evacuated and baked before filling.}
    \label{fig:alpha2}
\end{figure*}

\end{document}